\documentclass{PoS}
\usepackage{colordvi}
\usepackage{graphicx}
\usepackage{epsfig}
\usepackage{rotating}

\def\bra#1#2{\ifx#2\ket\langle#1\else\langle#1\vert\fi#2}
\def\ket#1{\vert#1\rangle}

\leftline{\parbox{3cm}{\large\rm HU-EP-08/39 LU-ITP 2008/02}
\vspace{1cm}}

\title{The Landau gauge lattice ghost propagator \\
       in stochastic perturbation theory}
\ShortTitle{The Landau gauge lattice ghost propagator in stochastic perturbation theory}

\author{
  Francesco~Di~Renzo \\
  Universit\`a di Parma \& INFN, Viale Usberti 7/A, I-43100 Parma, Italy \\
  E-mail: \email{francesco.direnzo@fis.unipr.it}
}
\author{
  \speaker{Ernst-Michael~Ilgenfritz}%
  \thanks{Supported by DFG via Forschergruppe Gitter-Hadronen-Ph\"anomenologie FOR 465.}\\
  Institut f\"ur Physik, Humboldt-Universit\"at zu Berlin, Newtonstr. 15, D-12489 Berlin, Germany \\
  Institut f\"ur Physik, Karl-Franzens-Universit\"at Graz, Universit\"atsplatz 5, A-8010 Graz, Austria \\
  E-mail: \email{ilgenfri@physik.hu-berlin.de}
}
\author{
  Holger~Perlt \\
  Institut f\"ur Theoretische Physik, Universit\"at Leipzig, PF 100 920, D-04009 Leipzig, Germany \\
  Institut f\"ur Theoretische Physik, Universit\"at Regensburg, Universit\"atsstr. 31, \\ D-93053 Regensburg, Germany \\
  E-mail: \email{Holger.Perlt@itp.uni-leipzig.de}
}
\author{
  Arwed~Schiller \\
  Institut f\"ur Theoretische Physik, Universit\"at Leipzig, PF 100 920, D-04009 Leipzig, Germany \\
  E-mail: \email{Arwed.Schiller@itp.uni-leipzig.de}
}
\author{
  Christian~Torrero \\
  Institut f\"ur Theoretische Physik, Universit\"at Regensburg, Universit\"atsstr. 31, \\ D-93053 Regensburg, Germany \\
  E-mail: \email{christian.torrero@physik.uni-regensburg.de}
}

\abstract{We present one- and two-loop results for the ghost propagator in 
Landau gauge calculated in Numerical Stochastic Perturbation Theory (NSPT).
The one-loop results are compared with available standard Lattice Perturbation 
Theory in the infinite-volume limit. We discuss in detail how to perform the 
different necessary limits in the NSPT approach and discuss a recipe to treat 
logarithmic terms by introducing ``finite-lattice logs''. We find agreement 
with the one-loop result from standard Lattice Perturbation Theory and estimate, 
from the non-logarithmic part of the ghost propagator in two-loop order, the 
unknown constant contribution to the ghost self-energy in the RI'-MOM scheme in 
Landau gauge. That constant vanishes within our numerical accuracy.}

\FullConference{The XXVI International Symposium on Lattice Field Theory\\
                 July 14 -- July 19, 2008\\
                 Williamsburg, VA, USA}

\begin{document}

\section{NSPT, Langevin equation, gauge fixing and all that}
\label{sec:basics}

Numerical Stochastic Perturbation 
Theory (for a review see Ref.~\cite{DiRenzo:2004ge}) is a powerful tool 
to study higher-loop contributions in Lattice Perturbation Theory (LPT). 
LPT is much more involved than perturbation theory in the continuum,
and thus only few results beyond one-loop level are available.
There have already been various applications of NSPT in the past: 
the average plaquette to very high orders in pure Yang-Mills theory 
to identify the gluon condensate~\cite{plaq}, the residual mass for lattice HQEF~\cite{Di Renzo:2004xn},
renormalization factors for bilinear quark operators~\cite{DiRenzo:2006wd}, 
renormalization factors related to the QCD pressure~\cite{DiRenzo:2006nh} etc.
Relatively new is the application of NSPT to gluon and ghost propagators 
in Yang-Mills theory~\cite{Ilgenfritz:2007qj,DiRenzo:2007qf}. 
Here we report on first steps towards an NSPT study of the ghost propagator in Landau gauge,
in particular at two-loop level.

It is known that the lattice Langevin equation with an additional 
running ``time'' $t$, beyond the four physical dimensions,
leads to a distribution of the gauge link fields according to 
the measure $\exp \left(- S_G[U]\right)$ in the limit $t\to \infty$.
Discretizing the time $t= n \tau$ and using the Euler scheme, the  
equation can be solved numerically by iteration: 
\begin{equation}
U_{x,\mu}(n+1; \eta)= {\rm{exp}}(-F_{x,\mu}[U,\eta])\; U_{x,\mu}(n; \eta)
\label{eq:update_step}
\end{equation}
with a force containing the gradient of $S_G$ and a Gaussian random noise $\eta$,
\begin{equation}
F_{x,\mu}[U, \eta]= {\rm{i}}(\tau \nabla_{x,\mu} S_G[U] + \sqrt{\tau} \, \eta_{x,\mu}) \, .
\label{eq:force}
\end{equation}
$\nabla_{x,\mu}$ is the left Lie derivative acting on gauge group-valued
variables while $S_G$ is Wilson's one-plaquette gauge action.

In NSPT one 
rescales $\varepsilon = \beta \tau$ and expands the 
link fields (and the force) in terms of the bare coupling constant $g \propto \beta^{-1/2}$:  
\begin{equation}
U_{x,\mu}(t; \eta) \to 1 + \sum_{l>0} \beta^{-l/2} U_{x,\mu}^{(l)}(t; \eta) \, .
\label{eq:U-expansion}
\end{equation}
Then the solution (\ref{eq:update_step}) transforms into a system of updates 
$U \to U^{\prime}$, one for each perturbative component $U^{(l)}$:
\begin{eqnarray}
U^{(1)'}  =  U^{(1)} - F^{(1)}\,, \quad
U^{(2)'}  =  U^{(2)} - F^{(2)} + \frac{1}{2} (F^{(1)})^2 - F^{(1)}U^{(1)}\,,
 \quad
\dots 
\label{eq:parallel_update}
\end{eqnarray}
The random noise $\eta$ is fed in only through $F^{(1)}$,
higher orders become stochastic by propagation of noise through the fields 
of lower order.

In terms of the (algebra-valued) gauge field variables
$A= \log U$, 
\begin{equation}
A_{x,\mu}(t; \eta) \to \sum_{l>0} \beta^{-l/2} A_{x,\mu}^{(l)}(t; \eta) \, , \quad  
A_{x,\mu}^{(l)}= T^a A_{x,\mu}^{a,(l)} \, ,
\label{eq:A-expansion}
\end{equation}
we are enforcing antihermiticity and tracelessness to all orders in $g$ by 
requiring
\begin{equation}
A^{(l)\dagger} = - A^{(l)} \, , \qquad {\rm{Tr}}A^{(l)}=0 \, .
\label{eq:antihermiticity}
\end{equation}
The  Landau gauge is achieved by iterative gauge transformations
using a perturbatively expanded version of the Fourier-accelerated
gauge-fixing method~\cite{Davies:1987vs} applied to each 50-th configuration
in the Langevin process. Only these are evaluated in order to control
the autocorrelations. Each Langevin update (\ref{eq:parallel_update}) 
is completed
by a stochastic gauge-fixing step and by subtracting zero modes of $A^{(l)}$
as described in Ref.~\cite{DiRenzo:2004ge}.

\section{The ghost propagator in NSPT and in standard LPT}
\label{sec:NSPT-ghost}

The continuum ghost propagator $G(q^2)$ in momentum space is defined as 
$G^{ab}(q)=\delta^{ab} G(q^2)$.
On the lattice it is obtained as the color trace
\begin{equation}
G(aq(k))= \frac{1}{N_c^2-1} G^{aa}(aq(k))=
\frac{1}{N_c^2-1} \left\langle {\rm{Tr}}~M^{-1}(k)\right\rangle_U
\label{eq:G-definition}
\end{equation}
as a function of the lattice momenta $a q_{\mu}(k)= 
2 \pi k_{\mu} a/L_{\mu}$ 
associated with plane waves $\ket{k}$ labelled by integers
$k_\mu=\left(-L_\mu/2,L_\mu/2\right]$.
In Landau gauge, the ghost propagator requires the computation of the inverse of the 
Faddeev-Popov (FP) operator
\begin{equation}
M= - \partial \cdot D(U) \, ,
\end{equation}
with $D(U)$ being the lattice covariant derivative and $\partial$ the
left lattice partial derivative.
$M^{-1}(k)$ in (\ref{eq:G-definition}) is the Fourier transform of the 
inverse FP operator. 

The perturbative expansion is based on the mapping
\begin{equation}
\{A_{x,\mu}^{(l)} \} \,\rightarrow\, \{M^{(l)} \} \,\rightarrow\, \{\left[M^{-1}\right]^{\!(l)}\} \, . 
\end{equation}
With an expansion of $M$ in terms of $M^{(l)}$ containing $A^{(l)}$, 
a recursive inversion is possible in coordinate space:
\begin{eqnarray}
\left[M^{~\!\!-1}\right]^{\!(0)} = \left[M^{~\!\!(0)}\right]^{\!-1} \,, \quad
\left[M^{~\!\!-1}\right]^{(l)} = -\left[M^{~\!0}\right]^{\!-1}\;\sum_{j=0}^{l-1} M^{~\!(l-j)}\;\left[M^{~\!\!-1}\right]^{\!(j)} \, .
\end{eqnarray}
The momentum-space ghost propagator at $n$-loop order is obtained from even orders $l = 2 n$ of $M^{-1}$ 
sandwiching its foregoing expansion between the plane-wave vectors:
\begin{equation}
 G^{~\!\!(n)}(aq(k)) = \langle k |\left[M^{~\!-1}\right]^{(l=2n)}| k \rangle \, . 
\end{equation}
Odd $l$ orders have to vanish numerically.
We discuss the results in terms of two forms of the dressing function for one and 
two loops:
\begin{equation}
J^{~\!\!(n)}(a q) = (a q )^2 \; G^{(n)}(aq(k)) \,, \quad \hat J^{~\!\!(n)}(\hat q) = \hat q^2 \; G^{(n)}(aq(k)) . 
\end{equation}
Here we use the standard notation for hat-variables, {\em e.g.}
\begin{equation}
\hat q_{\mu}(k_{\mu}) = 
\frac{2}{a} \sin\left(\frac{\pi k_{\mu}}{L_{\mu}}\right) = 
\frac{2}{a} \sin\left(\frac{ a q_\mu}{2}\right) .
\end{equation}

In standard LPT, loop contributions are calculated in the infinite volume and $a \to 0$ limit.
In this limit the two dressing functions coincide. 
The renormalization of the dressing function is performed in the RI'-MOM scheme:
\begin{equation}
J^{~\!\!\rm RI'}( q, \mu , \alpha_{\rm RI'}) = \frac{J(a, q, \alpha_{\rm RI'})}{Z_{\rm gh}(a,\mu,\alpha_{\rm RI'})} 
\end{equation}
with the renormalization condition
\begin{equation}
J^{~\!\!\rm RI'}( q, \mu , \alpha_{\rm RI'})|_{q^2=\mu^2} = 1 \,. 
\end{equation}
Restricting ourselves to two-loop order, we have {\it e.g.}
\begin{equation}
J(a,q, \alpha_{\rm RI'}) = 1 + \sum_{i=1}^{2} \alpha_{\rm RI'}^i \,\sum_{k=0}^i\, z^{~\!\rm RI'}_{i,k}\,
\left(\frac{1}{2}\log(aq)^2\right)^{\!\!k} \, .
\end{equation}
Only the leading coefficients $z^{~\!\rm RI'}_{i,i}$ are entirely calculable in continuum 
perturbation theory (PT): $z^{~\!\rm  RI'}_{1,1}= - 3N_c/2$, $z^{~\!\rm RI'}_{2,2}= - 35N_c^{~\!\!2}/8$.
The non-leading coefficients 
$z^{~\!\rm RI'}_{i,k}|_{i>k>0}$ are 
only partly known from PT: $z^{~\!RI'}_{2,1}= \left(- \frac{271}{24} + \frac{35}{6} \, z^{~\!RI'}_{1,0}  \right)$,
the $z^{~\!\rm RI'}_{i,0}$ have to be calculated in LPT.
For example, entering $z^{~\!\rm RI'}_{2,1}$ is 
$z^{~\!\rm  RI'}_{1,0}=13.8257$, known from one-loop LPT~\cite{Kawai:1980ja}, while
$z^{~\! \rm  RI'}_{2,0}$ is unknown.

 From the relation~\cite{Hasenfratz:1980kn} 
$
\alpha_{\rm RI'}= \alpha_0 + \left(-(22/3) \, N_c \log (a\mu) + 73.9355 \right) \, \alpha_0^2 + \dots \, , 
$
with the bare coupling $\alpha_0={N_c}/{(8 \pi^2 \beta)}$, we get for the two-loop 
dressing function:
\begin{eqnarray}
J^{\rm{2-loop}}(a,q, \beta) = 1  +  \frac{1}{\beta} \left(J_{1,1}  \log(aq)^2 + J_{1,0} \right) 
 +  \frac{1}{\beta^2}\left(J_{2,2} \log^2(aq)^2 +J_{2,1}  \log(aq)^2 + J_{2,0}\right) 
\end{eqnarray}
with
\begin{eqnarray}
  J_{1,1}  =  -0.0854897 \, ,  \quad  J_{1,0}  =  0.525314 \, , \quad
  J_{2,2}  =   0.0215195 \, ,  \quad  J_{2,1}  = -0.358423 
\end{eqnarray}
and the unknown finite two-loop finite 
constant $J_{2,0}$ or $z^{~\!\rm RI'}_{2,0} \, ,$
\begin{equation}
J_{2,0}=1.47572 + 0.00144365 \, z^{\rm RI'}_{2,0} \, .
\label{eq:J20}
\end{equation}

\section{Results}
\label{sec:results}

The aim of this first investigation of the ghost propagator in NSPT was 
the confirmation of the known $J_{1,0}$, 
and a prediction of the unknown $J_{2,0}$.
We concentrate ourselves on an analysis of $\hat{J}^{~\!(n)}(\hat q)$.

\begin{figure}[!htb]
\begin{center}
\begin{tabular}{cc}
\includegraphics[scale=0.59,clip=true] {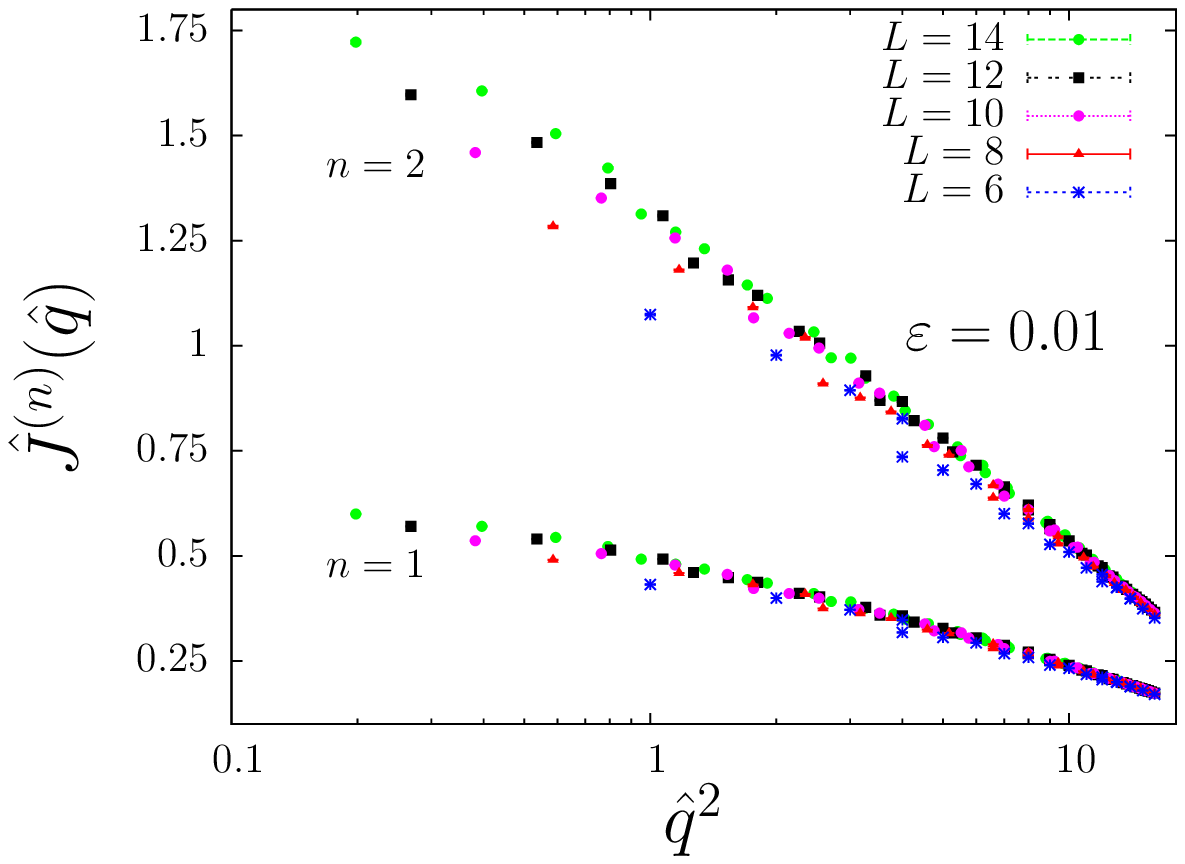}&
\includegraphics[scale=0.59,clip=true] {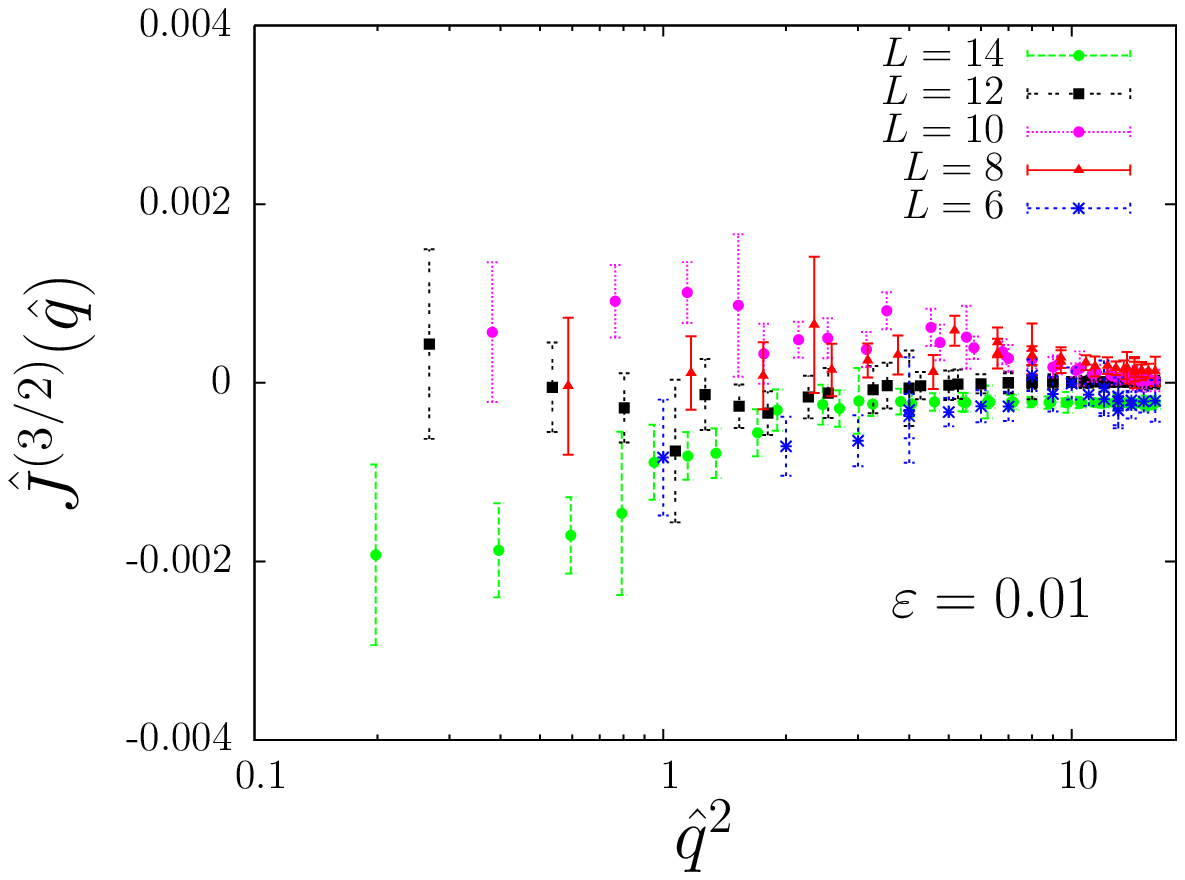} \\
\end{tabular}
\end{center}
\vspace{-5mm}
\caption{Measured ghost dressing function $\hat{J}(\hat q)$ vs. $\hat q^2$ 
for all inequivalent lattice momentum 4-tuples $(k_1,k_2,k_3,k_4)$ -
see (2.2) -
near the diagonal ones
for lattice sizes $L=6,\dots,14$ and for the time step $\varepsilon=0.01$.
Left: The one-loop ($\propto \beta^{-1}$) and 
two-loop ($\propto \beta^{-2}$) contributions, right: the vanishing 
contribution $\propto \beta^{-3/2}$.}
\label{fig:one_and_two_loop}
\end{figure}

As an example of the measured ghost propagator
we show the one- and two-loop results $\hat{J}^{~\!(1)}$ and $\hat{J}^{~\!(2)}$ 
for the dressing function in Fig.~\ref{fig:one_and_two_loop} 
together with $\hat{J}^{~\!(n=3/2)}$ that is bound to vanish.

\subsection{The limits to be taken}
\label{subsec:limits}

\begin{itemize}
\item The limit $\varepsilon \to 0$: 
We solved the Langevin equations for different step sizes 
$\varepsilon= 0.07, \ldots, 0.01$ and obtained the Langevin result for
each chosen momentum set of the propagator at fixed lattice size $L$ and 
$\varepsilon=0$ by extrapolation as
shown in Fig.~\ref{fig:epsilon_extrapolation}.
\begin{figure}[!htb]
\begin{center}
\begin{tabular}{cc}
\includegraphics[scale=0.59,clip=true]{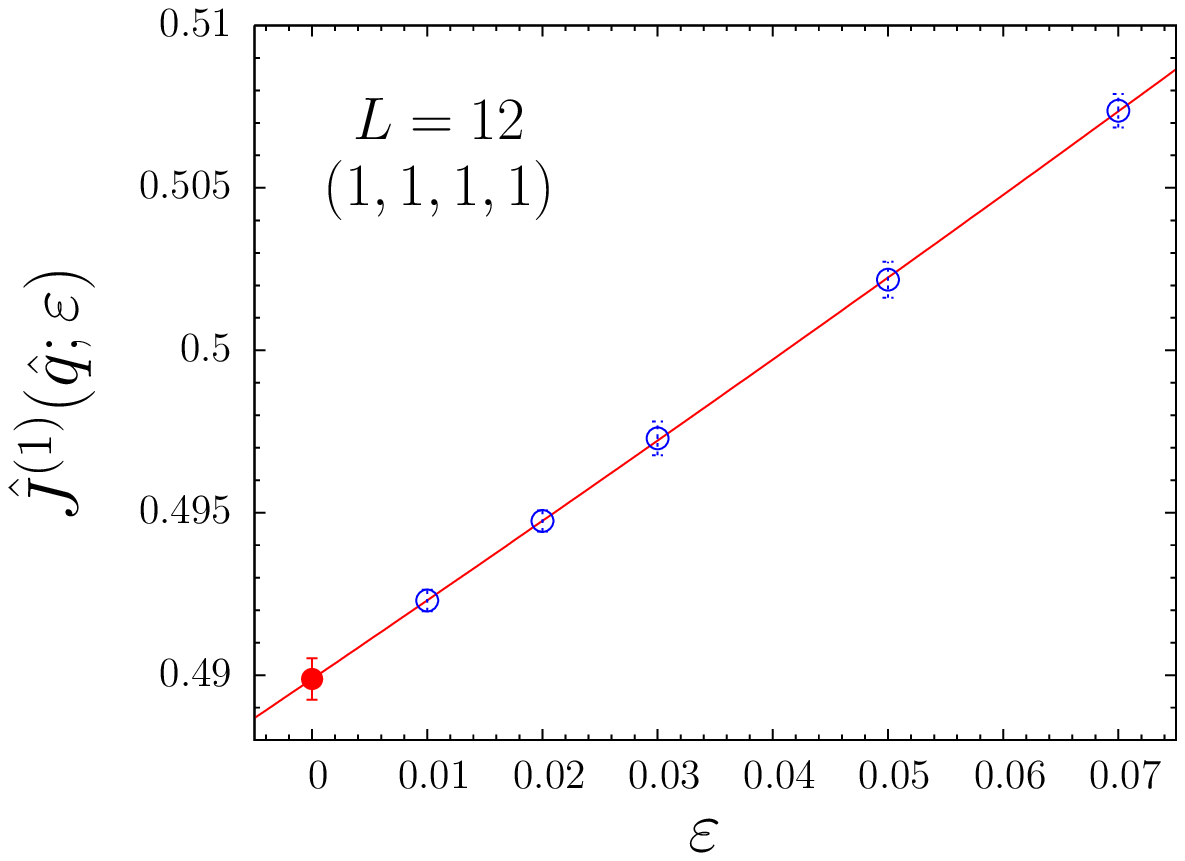}&
\includegraphics[scale=0.59,clip=true]{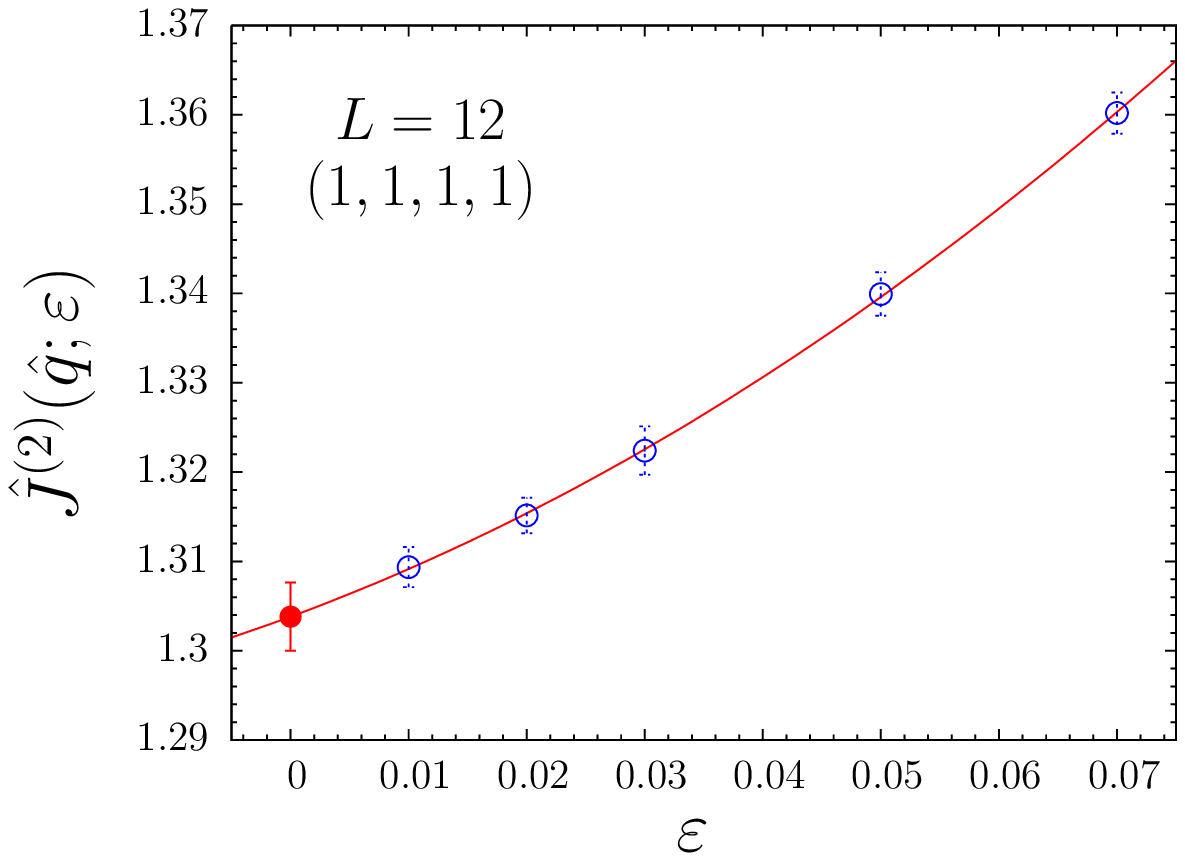}\\
\end{tabular}
\end{center}
\vspace{-5mm}
\caption{Linear plus quadratic correction extrapolation to $\varepsilon=0$ 
of the one-loop (left) and two-loop (right) ghost dressing function for
the momentum tuple  $(1,1,1,1)$ on a lattice of size $12^4$.}
\label{fig:epsilon_extrapolation}
\end{figure}

\item The limits $L \to \infty$ and $a \to 0$:
In order to make contact with standard LPT both limits have to be performed.
To extract the non-logarithmic constants in those limits we make the following ansatz
for the dressing function taking into account hypercubic symmetry
(one-loop example; here we use the standard notation for hypercubic invariants)
\begin{eqnarray}
 &&  {\hat J}^{~\!(1)}(\hat q)= J_{1,1} `` \log \hat q^2 " + {{\hat J}_{1,0;L}} (\hat q)\,,
\\
&& \hat J_{1,0;~\!L}(\hat{q}) =\hat J_{1,0;~\!L}
+ c_{1} \, \hat{q}^2
+ c_{2}\,  \frac{\hat{q}^{~\!\!4}}{\hat{q}^{~\!\!2}}
+ c_{3} \, \hat{q}^{~\!\!4}
+ c_{4} \, (\hat{q}^{~\!\!2})^2
+ c_{5}\,  \frac{\hat{q}^{~\!\!6}}{\hat{q}^{~\!\!2}}
+ c_{6} \, (\hat{q}^{~\!\!2})^3
+ \cdots
  \label{eq:improvedfitghost}
\end{eqnarray}
The problem arising here is how to represent -- on finite lattices -- 
the logs that appear in the $L \to \infty$ regime.
Our proposal here is to replace the divergent lattice integrals, that give rise
to the logarithms, by finite lattice sums and use these expressions in the 
fits at fixed $L$.
\end{itemize}

\subsection{Handling the lattice logs encountered}
\label{subsec:logs}

\begin{figure}[!htb]
\begin{center}
\includegraphics[scale=0.58,clip=true]{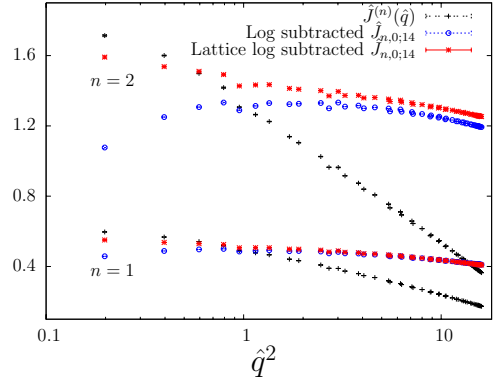}
\end{center}
\vspace{-5mm}
\caption{Original and remaining ``non-logarithmic'' contributions 
to $\hat{J}$ using logarithms and lattice logarithms at one-loop  
and two-loop level as function of $\hat q^2$ for a lattice $14^4$.}
\label{fig:subtracted}
\end{figure}

We illustrate the procedure by the example of a typical one-loop divergent integral
\begin{equation}
A(a q) = (4 \pi)^2  \,   \int_{-\pi/a}^{\pi/a} \frac{d^4 k }{(2 \pi)^4}\frac{1}{\hat k^2 \widehat{(k+q)}^2} \, .
\label{eq:Adef1}
\end{equation}
In the limit $a q\to 0$ \cite{Luscher:1995zz} one gets
\begin{equation}
A(a q) = - \log (a q)^2 + a_1 \, , \qquad a_1=2+F_0-\gamma_E=5.79201 \, .
\label{eq:Adef2}
\end{equation}
On a lattice with finite $L$ we calculate the corresponding lattice sums:
\begin{equation}
A(i^{~\!q},L) = 
  \frac{1}{L^4}
\sum_{i_1,i_2,i_3,i_4} 
\frac{1}{\left[\sum_{\mu=1}^4  \sin^2 \left(\frac{\pi}{L} i_\mu \right)\right] 
\left[\sum_{\nu=1}^4  \sin^2 \left( \frac{\pi} {L} (i_\nu -i_\nu^{~\!q} ) \right) \right] } 
\end{equation}
with
$a k_\mu= \frac{2 \pi i_{~\!\!\!\mu}}{L}$, $a q_\mu= \frac{2 \pi i_{~\!\!\!\mu}^{~\!\!q}}{L}$,
$\{i_{\mu},i_{\mu}^{~\!\!q} \}\in \left(-\frac{L}{2}, \frac{L}{2}\right]$.
This leads -- for each $L$ -- to the replacement:
\begin{eqnarray}
J_{1,0} \log (a q)^2 \to 2~J_{1,0} \, (A(i^{~\!q} ,L)-a_1) \ . 
\end{eqnarray}
This also results in a reshuffling of irrelevant terms. 
The result is a flattening of the data with the log-terms subtracted
(see Fig.~\ref{fig:subtracted}).
This then allows to extract the $V \to \infty$ limit fitting the remaining non-logarithmic data
(at present no momentum cuts on the data are used)
with the ansatz (\ref{eq:improvedfitghost}).
In a similar spirit, a log-squared behavior in a two-loop contribution is modeled by using
the following expression as a discretized version of~\cite{Luscher:1995zz}
\begin{eqnarray}
E(a q)=(4 \pi)^4 \, \int_{-\pi/a}^{\pi/a} \frac{d^4 k }{(2 \pi)^4}\frac{1}{\hat k^2 \widehat{(k+q)}^2}  A(a k)
{\to} \frac{1}{2} \log^2 (aq)^2 - (a_1+1) \log (aq)^2 + 28.0086 
\end{eqnarray}
where $A(a k)$ and $a_1$ are defined in (\ref{eq:Adef1}) and (\ref{eq:Adef2}).

\subsection{Results based on the outlined fitting procedure}
\label{subsec:final}

\begin{figure}[!htb]
\begin{center}
\begin{tabular}{cc}
\includegraphics[scale=0.57,clip=true]{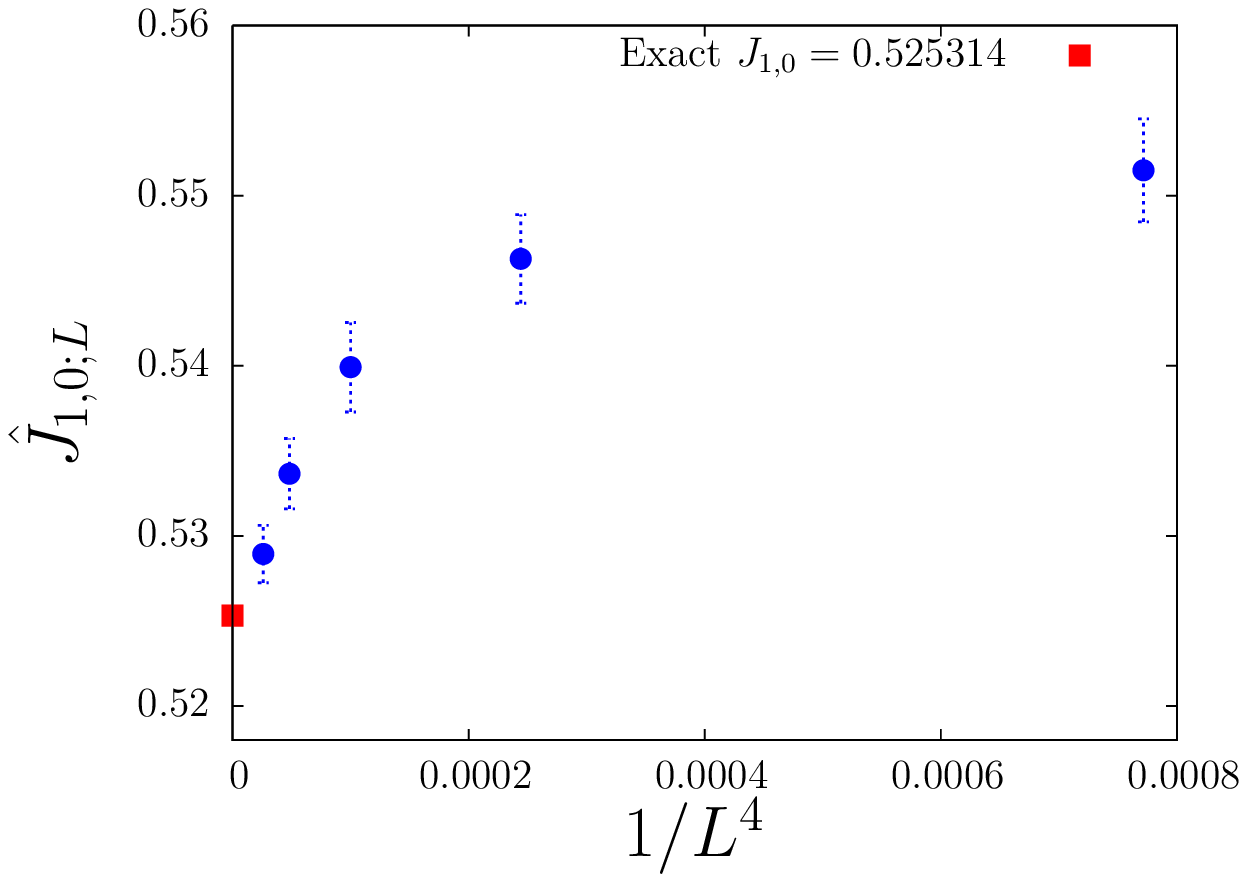}&
\includegraphics[scale=0.57,clip=true]{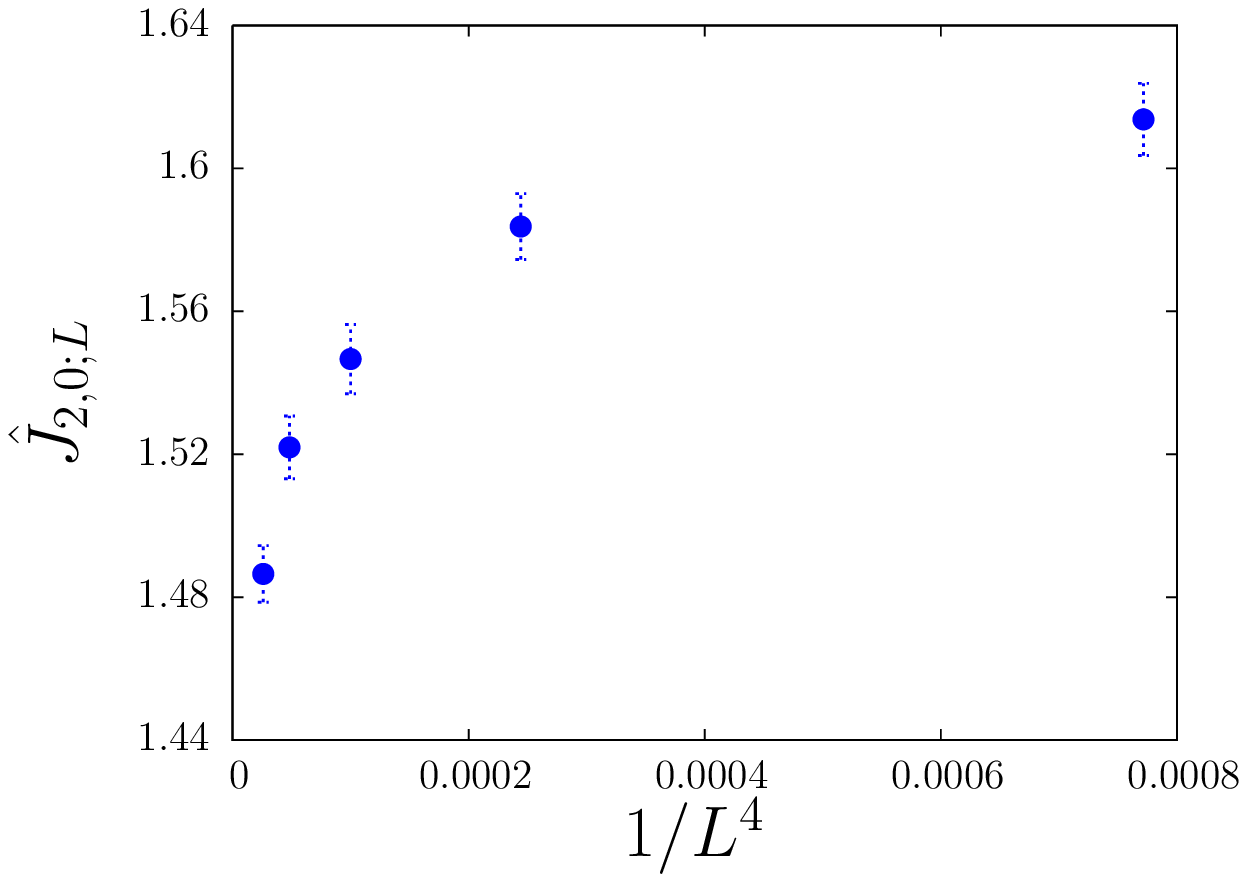}
\end{tabular}
\end{center}
\vspace{-5mm}
\caption{The $V\to \infty$ limit of the constant $\hat{J}_{n,0;~\!\!L}$.}
\label{fig:extrapolation}
\end{figure}

The results for $\hat{J}_{1,0;~\!L}$ and $\hat{J}_{2,0;~\!L}$ as function of $1/L^4$ 
are shown in Fig.~\ref{fig:extrapolation}. A linear fit for $L=10,12,14$ leads to 
the one-loop result
\begin{equation}
\hat{J}_{1,0}^{~\!\rm Fit} = 0.5255(24) 
\end{equation}
in agreement with the expectations.
A linear fit as in the one-loop case would lead to a preliminary two-loop
value  $\hat{J}_{2,0}^{~\!\rm Fit} = 1.47(2)$.
This results in the non-logarithmic contribution $z^{~\!\rm RI'}_{2,0}$
to the two-loop ghost self-energy
in the RI'-MOM scheme in Landau gauge being compatible with zero.

\section{Summary}
\label{sec:summary}
\begin{itemize}
\item We have performed the first two-loop calculation of the 
lattice ghost propagator in Landau gauge.
\item The one-loop constant $J_{1,0}$ agrees with the known 
$V \to \infty$ result.
\item The two-loop constant $J_{2,0}$ has been estimated for
the first time.
\item A detailed analysis of all necessary limits has been performed.
\item A proposal about how to mimic the usual logarithmic terms on finite 
lattices is made. An alternative procedure outlined 
in Ref.~\cite{DiRenzo:2007qf} 
is under development.
\item A detailed comparison for a finite volume and a set of lattice 
momenta  with Monte Carlo data would be desirable in order to separate 
out the nonperturbative effects on the ghost propagator.
\end{itemize}

\section*{Acknowledgements}
Part of this work is supported by DFG under contract FOR 465 
(Forschergruppe Gitter-Hadronen Ph\"anomenologie).
E.-M.~I. is grateful to the Karl-Franzens-Universit\"at Graz for the
guest position he holds while this paper is written up.


\begin{thebibliography}{99}

\bibitem{DiRenzo:2004ge}
F.~Di~Renzo and L.~Scorzato, 
JHEP {\bf 10} (2004) 073 
[\href{http://xxx.lanl.gov/abs/hep-lat/0410010}{\tt arXiv:hep-lat/0410010}].

\bibitem{plaq}
F.~Di Renzo, E.~Onofri and G.~Marchesini,
Nucl.\ Phys.\  B {\bf 457} (1995) 202. \\
P.~E.~L.~Rakow, 
 \pos{PoS(LAT2005)284} 
[\href{http://xxx.lanl.gov/abs/hep-lat/0510046}{\tt arXiv:hep-lat/0510046}].

\bibitem{Di Renzo:2004xn}
F.~Di Renzo and L.~Scorzato,
JHEP {\bf 0411}, 036 (2004)

\bibitem{DiRenzo:2006wd}
F.~Di~Renzo, V.~Miccio, L.~Scorzato, and C.~Torrero, 
Eur.~Phys.~J. {\bf C51} (2007) 645 
[\href{http://xxx.lanl.gov/abs/hep-lat/0611013}{\tt arXiv:hep-lat/0611013}].

\bibitem{DiRenzo:2006nh}
F.~Di~Renzo, M.~Laine, V.~Miccio, Y.~Schr\"oder, and C.~Torrero, 
JHEP {\bf 07} (2006) 026 
[\href{http://xxx.lanl.gov/abs/hep-ph/0605042}{\tt arXiv:hep-ph/0605042}].

\bibitem{Ilgenfritz:2007qj}
E.-M.~Ilgenfritz, H.~Perlt, and A.~Schiller, 
 \pos{PoS(LATTICE 2007)251} 
[\href{http://arxiv.org/abs/0710.0560}{arXiv:0710.0560[hep-lat]}].

\bibitem{DiRenzo:2007qf}
F.~Di~Renzo, L.~Scorzato, and C.~Torrero, 
 \pos{PoS(LATTICE 2007)240} 
[\href{http://arxiv.org/abs/0710.0552}{arXiv:0710.0552[hep-lat]}].

\bibitem{Davies:1987vs}
C.~T.~H.~Davies et al.,
Phys.~Rev.~{\bf D 37} (1988) 1581. 

\bibitem{Kawai:1980ja}
  H.~Kawai, R.~Nakayama and K.~Seo,
  Nucl.\ Phys.\  B {\bf 189} (1981) 40.

\bibitem{Hasenfratz:1980kn}
  A.~Hasenfratz and P.~Hasenfratz,
  Phys.\ Lett.\  B {\bf 93} (1980) 165.

\bibitem{Luscher:1995zz}
M.~L\"uscher and P.~Weisz, 
Nucl.~Phys.~{\bf B 445} (1995) 429 
[\href{http://xxx.lanl.gov/abs/hep-lat/9502017}{\tt arXiv:hep-lat/9502017}].

\end{thebibliography}
\end{document}